\documentclass[krantz1,ChapterTOCs]{krantz} 
\usepackage{fixltx2e,fix-cm}
\usepackage{amssymb}
\usepackage{amsmath}
\usepackage{graphicx}
\usepackage{subfigure}
\usepackage{makeidx}
\usepackage{multicol}

\makeindex

\begin{document}







\tableofcontents

\chapterauthor{John M. Abowd and Michael B. Hawes}{U.S. Census Bureau}
\chapter{$21^{st}$ Century Statistical Disclosure Limitation: Motivations and Challenges}

Over the coming decade, national statistical offices will likely undertake a re-engineering of their data confidentiality programs comparable in magnitude to the transformation of statistical disclosure limitation (SDL) that began in the 1970s. Fellegi \cite{fellegi:1972:doi:10.1080/01621459.1972.10481199} and Delenius \cite{delenius} ushered in a principled and scientific approach to SDL that fundamentally reshaped how statistical agencies assessed and controlled disclosure risk in their public data releases. Over the subsequent decades, agencies continued to improve and strengthen their implementations of SDL, but these changes have largely been incremental adjustments and extensions to approaches pioneered in the 1970s and 1990s. Today, advances in computing power, the development of powerful optimization algorithms, and the proliferation of rich, third-party data have contributed to a data protection landscape that renders the widely used SDL methods of the last several decades increasingly vulnerable. Modernization of SDL for the $21^{st}$ Century is not going to be easy nor will it be uncontroversial. Not only will it require statistical agencies to rethink their entire approach to SDL and how it fits within the broader data life cycle, but it will also require agencies and data users alike to make difficult decisions about the content and form of official statistics and how data users can access them.

This chapter examines the motivations and imperatives for modernizing how agencies approach SDL for official statistics. It discusses the implications of this modernization on agencies' broader data governance and decision-making, and it identifies challenges that agencies will likely face along the way. In conclusion, we propose some principles and best practices that we believe can help guide agencies in navigating the transformation of their confidentiality programs.

\section{Motivations}
The central mission of any national statistics office is, by definition, the production of official statistics. The production of accurate and objective statistics about a nation's people and economy is ``an indispensable element in the information system of a democratic society'' and reflects ``citizens' entitlement to public information'' (OECD 2005, Principle 1) \cite{OECD}. Used appropriately, official statistics serve as a critical input to myriad worthwhile activities, including governmental decision-making, private investment, and scientific research, to name just a few. But the detailed information about a nation's people, businesses, health, safety, and organizations that is necessary to create this rich societal portrait, if misused, could be profoundly harmful to the data subjects from (or about) which the information was collected. To encourage the social benefits that accrue from access to statistical information, nations often enable their statistical offices to access and publish summary data from nonstatistical (or administrative) agencies often encouraging commingling of these administrative data with information collected directly by the statistical agency. To prevent the misuse of information produced for official statistical purposes, nations often create two legislative programs. First, they enact strong confidentiality laws that limit access to, and prohibit publication of, identifiable information collected for statistical purposes. Second, they establish regulatory control over the types of information collected and the allowable uses throughout the government, including by statistical offices.

Thus, in pursuit of their mission to produce official statistics, agencies face countervailing obligations to develop statistical summaries that comply with allowable data usage and to ensure that data subjects' identifiable information is properly safeguarded in the agencies' publications. To these ends, agencies perform project reviews and disclosure risk assessments on the statistics they intend to publish. They limit statistical analyses to approved projects and apply SDL methods, as necessary, to mitigate the risk of disclosure.

\subsection{SDL in historical perspective}
The goal of SDL is to produce detailed and accurate summaries of the characteristics of a population without revealing the characteristics or the identity associated with those characteristics of any of the specific individuals/entities within that population. The complexity of SDL methods varies widely, but even the simplest SDL mechanisms like aggregation or redaction can be effective depending on the context. In general, however, the more granular the aggregate statistics or the more detailed the individual-level records an agency seeks to publish, the easier it will be to identify specific data subjects and the less effective those simple SDL mechanisms will be at reducing disclosure risk. Over the decades, as official statistics have increased in scale, scope, and detail, agencies have typically had to rely on increasingly complex SDL mechanisms to manage this risk.

Historically, most of the confidentiality laws under which statistical agencies operate and many of the SDL frameworks that agencies have relied upon to meet those confidentiality obligations have rested on the assumption that once identifiable data are properly treated with SDL, the resulting data are ``de-identified'' and carry no risk of disclosure. And, again historically, the leading technical handbooks have reinforced this idea, albeit with appropriate caveats about risk limitation \cite{Duncan:et:al:2011} or control \cite{Willenborg:deWaal:2000}. Neither textbook provides a method of accounting for global disclosure risk from multiple releases, tabular and microdata, from the same confidential sources. 

This binary divide between identifiable data that agencies must protect and de-identified data that they can publicly release with no further consideration was convenient. It made the legal confidentiality frameworks for statistical data appear clear-cut and absolute, and it simplified agency governance of the disclosure review process, leading curators to approach SDL for each data product individually on a ``release and forget'' model. Unfortunately, the underlying assumption that identifiable data can, through proper application of SDL, be de-identified is false \cite{garfinkel:et:al:2023}. The release of any statistic derived from a confidential source always carries some incremental risk of disclosure of identifiable information. This unfortunate truth has been acknowledged in principle since the 1970s \cite{FCSM:SWP2:1978} and was proven mathematically in 2003 \cite{dinur:nissim:2003:10.1145/773153.773173}.

Addressing the reality of disclosure risk in the publication of official statistics has not been easy. While most confidentiality laws still maintain the false dichotomy between identifiable and de-identified data, most statistical agencies have, over the past fifty years, transitioned in varying degrees towards a disclosure risk management framework.

Adopting a disclosure risk management framework for official statistics in the context of traditional SDL methods and within agencies' data production cycles, is far from straightforward. Statistical agencies often publish multiple data products in sequence from the same underlying confidential data. Most traditional SDL methods, however, are unable to objectively quantify disclosure risk across multiple data releases, and those that do fail if they are not applied without exception to every statistic released from the confidential source. Furthermore, the operational realities of agencies' rapid and tight data production schedules favor rules-based, release-by-release application of SDL in a short window just prior to publication.

Consequently, for much of the last half-century, most statistical agencies have relied on largely subjective assessments of the overall disclosure risks associated with different types of data products and the establishment of rules and manuals for how SDL should be applied in each context. These approaches favored the adoption of relatively simple rounding and/or suppression rules for most aggregate, tabular statistics, and more sophisticated suppression, coarsening, or noise injection rules for microdata releases.

Some statistical agencies, like the U.S. Census Bureau, buttressed this approach to disclosure risk management, by conducting periodic re-identification studies on specific data products and by using the results of those studies to adjust or augment the SDL rules to be followed for subsequent data releases. The problem with these re-identification studies, however, is that they rely heavily on specific assumptions about what types of attack are feasible and what resources (e.g., computing infrastructure and external information) an attacker might have at their disposal. As such, these studies provide, at best, a limited snapshot of the overall disclosure risk of a data product, at a particular moment in time. 

Cynthia Dwork's work at Microsoft Research \cite{dwork:2006:10.1007/11787006_1} opened the door to solving the problem of controlling cumulative disclosure risk by introducing the first formally private SDL framework---differential privacy. She provided a concise, plain language summary of formal privacy: ``we will first define our privacy goals and then explore what utility can be achieved given that the privacy goals will be satisfied'' (p. 2). Formal privacy systems provide a mathematical definition of a confidentiality breach. SDL methods that provably satisfy the definition are called implementing mechanisms. For example, the seminal paper in differential privacy \cite{dwork:et:al:2006:10.1007/11681878_14} defines a confidentiality breach as an increase of more than $\epsilon$ in the worst-case inference by a user of whether an entity was included or excluded from a statistical tabulation. This framework is now called pure or $\epsilon$--differential privacy. They prove that adding double exponential noise with a scale parameter that depends on $\epsilon$ and the worst-case feasible change in the statistical tabulation satisfies the definition of pure differential privacy. We now call this implementing pure differential privacy via the Laplace mechanism. When we reference formal privacy methods in this paper, we mean SDL frameworks that provide a mathematical definition of a confidentiality breach and at least one feasible implementing mechanism satisfying that definition.

Another consequence of the convenient fiction of de-identification and subsequent rules-based approaches to disclosure risk management is that decision-making about SDL was widely perceived, both publicly and by agency leaders, as the purview of technical SDL practitioners within the agency. As such, there has historically been little public discussion or debate about how much disclosure risk is acceptable or about the impact of SDL methods on data utility.\footnote{This latter point being further exacerbated in the context of some SDL methods (e.g., for some swapping and noise infusion mechanisms) by the conventional practice of keeping the methods' parameters and impact on data accuracy confidential to avoid undermining the mechanisms.} While the design and application of effective SDL is a highly technical discipline, the decision-making underlying disclosure risk assessment and mitigation has major public policy implications. We discuss in this article how the deliberations about disclosure risk benefit greatly from enterprise-level coordination of the framework and parameters for agency-wide SDL, coordination with internal subject-matter experts, and guidance informed by external stakeholder feedback.

\subsection{Disclosure risk in the $21^{st}$ Century}
None of the limitations associated with agencies' traditional approaches to SDL discussed above would really have come as a surprise to experienced SDL practitioners over the past few decades. Fellegi himself acknowledged the vulnerability of rules-based approaches to SDL when applied across disparate tables or data releases \cite{fellegi:1972:doi:10.1080/01621459.1972.10481199}. But agencies generally considered the benefits and simplicity of rules-based approaches by data type on a release-by-release basis sufficient to justify the increased disclosure risk. And those re-identification studies that were performed largely supported this conclusion. 

Technological developments over the last decade, however, have brought the limitations of traditional SDL approaches into much clearer focus and have demonstrated the fundamental problem with data type-specific rules-based SDL. Historical disclosure risk assessments focused largely on simple subtraction attacks against tabular data and record linkage-based re-identification attacks on microdata products. This led to the almost ubiquitous practice of agencies employing one set of SDL rules for tabular data products and a second set, often but not always more stringent, of SDL rules for microdata releases, where the underlying disclosure risk was usually deemed higher.

The availability of massive cloud computing platforms and powerful optimization algorithms have made database reconstruction attacks, first predicted by Dinur and Nissim \cite{dinur:nissim:2003:10.1145/773153.773173}, feasible and widely accessible, opening up a new vector of attack on official statistics not considered by traditional disclosure risk assessments. The Census Bureau's simulated attack on the published 2010 Census tabulations demonstrated that it was possible to generate a high quality reconstructed record image (for over 65\% of census blocks, a perfect image) of the microdata used to generate the tables. Those microdata were not considered releasable under the SDL rules in place in 2010 \cite{abowd:hawes:2023}. This simulated attack demonstrates, at least for large scale tabular data releases, that SDL for aggregate tabular data cannot be considered separately from SDL for microdata. Given enough data, aggregate statistics can be equivalent to microdata.

Traditional disclosure risk assessments underestimated the disclosure risks associated with tabular data because their underlying assumptions did not include the possibility of a reconstruction-based attack vector. Similarly, it would be naive for agencies to assume that reconstruction-abetted re-identification studies would not also underestimate overall disclosure risk in the face of emerging or future attack vectors. Inference and membership attacks, for example, are emerging vectors that agencies should consider. Nissim \cite{nissim:2021} provides a framework to relate many of these attacks to the legal requirements of agency confidentiality laws.

If there is one lesson to be learned from the emergence of database reconstruction as a viable vector of attack on official statistics it is that disclosure risk can only be effectively managed if the incremental disclosure risk across multiple releases (be they statistics, tables, or microdata products) is quantified and controlled. It is this composability across multiple data releases that most traditional SDL methods lack and why they will be increasingly likely to fail over time. SDL methods based on formal privacy frameworks, however, offer agencies the ability to precisely quantify the incremental disclosure risk of each statistic they publish because their mathematical definitions provably compose. This composition property means that such frameworks provide an accounting system that can be used across multiple data products to quantify incremental disclosure risk. This is not to say that the adoption of formal privacy will eliminate the risk of disclosure in official statistics. On the contrary, formally private SDL can be implemented anywhere along the spectrum from complete disclosure risk aversion to complete risk acceptance, depending on the decisions made by the agency. It is the precise accounting of incremental risk that informs and enables that decision-making.

Furthermore, while traditional approaches to SDL were, by convention or convenience, largely technocratic and opaque to data users, the transparency possible under formally private SDL approaches should not be undervalued. One of the guiding principles of statistics as a discipline is the importance of transparency about known limitations of one's data or analysis \cite{asa:2022}. The selection and implementation of any SDL method will necessarily impact the fitness-for-use of the resulting data. Statistical agencies should be transparent with their data users about those impacts.

\section{Considerations}
Moving SDL choices out of the backroom and into public and professional debates about the appropriate trade-offs requires acknowledging that global confidentiality risk management involves three choices, not two.

\subsection{The triple trade-off of official statistics}
Public debates about SDL generally, and about formal privacy mechanisms in particular, tend to focus on the centrality of the trade-off between privacy and accuracy/utility \cite{hotzetal}. This trade-off is undeniably important and it forms the basis for interpreting the privacy-loss accounting (and corresponding fitness-for-use) of formally private SDL implementations. But focusing only on the trade-off between privacy and accuracy simultaneously oversimplifies and complicates decision-making. Accuracy or data utility in the context of SDL is not a one-dimensional characteristic---there is no single, universal measure for assessing data utility. 
More fundamentally, approaching SDL decision-making from the perspective of privacy versus accuracy complicates another important dimension: the quantity of statistics to be published. The same privacy-loss budget can be distributed over a few very accurate statistics or many less-accurate ones. The utility of either choice depends upon the applications the statistics are intended to support.

Dinur and Nissim \cite{dinur:nissim:2003:10.1145/773153.773173} demonstrated that publishing too many accurate statistics will undermine confidentiality. Statistical agencies could, theoretically, release highly accurate statistics with practically zero risk of disclosure if they only published a small number of statistics. Consequently, rather than focusing on a trade-off between privacy and accuracy for a fixed-dimension statistical output, public debate and agency decision-making about SDL in the context of official statistics should focus on a broader triple trade-off:  privacy vs. accuracy vs. availability \cite{hawes2021}. In a triple trade-off, maximizing on any single dimension is easy, and maximizing on any two dimensions is possible, but only at profound impact to the third. Thus, agencies could publish large amounts of accurate data, but with substantial risk of violating confidentiality. Or, they could publish large amounts of data with strong confidentiality protections, but only by significantly decreasing the utility of the resulting data for at least some applications. Or, they could conduct independent surveys to estimate different statistics, thus ensuring strong confidentiality but reducing accuracy if the same operational costs are split between the surveys. Lastly, they could publish data that are highly accurate and highly protected, but only by massively curtailing the quantity of statistics released. Agencies that use properly implemented primary/complementary suppression \cite{cox:1995}, like the economic census data products released by the U.S. Census Bureau, already impose massive curtailment. 
The quantity dimension is also a question of ``utility for whom?'' Although research along these dimensions is still in its infancy, under formal privacy frameworks, budgets could be allocated unequally across queries, which means different privacy guarantees for different features of the data for some inferences \cite{kifer:et:al:2022}.
Hard choices about which dimensions of the data to favor for more granular tables inevitably restrict the utility of the releases.

For agencies to navigate the countervailing objectives of producing statistics that are accurate and relevant for societal decision-making while protecting confidentiality, concessions must be made along all three dimensions: privacy, accuracy and quantity. This interplay between all three dimensions underscores the importance of involving senior leadership in decision-making about SDL, rather than treating it as the exclusive province of the SDL practitioners charged with designing and implementing the protections.

\subsection{Need for consistency in evaluation of acceptable risk}
The evolution of SDL from purported de-identification of data to disclosure risk management centers on the presumption of an acceptable level of disclosure risk. In the context of confidentiality laws that do not acknowledge the inherent disclosure risk of any data release, there is no objective standard for how much risk is acceptable. Agencies must make these determinations for themselves balancing the public value of the data to be released against their interpretation of their legal and ethical obligations to protect confidentiality and their own perceived tolerance or aversion to a potential disclosure occurring. Because the potential harm from disclosure varies greatly by the type of data involved (e.g., financial, demographic, and health data), ensuring comparable and consistent evaluation and mitigation of disclosure risk across data products requires enterprise-wide governance of how disclosure risk assessments are made.  

\section{Challenges}
Navigating the triple trade-off of official statistics, as discussed above, requires agencies to make important decisions about the scope and content of their data products, weighing data users' needs and the public value of official statistics against the legal and ethical imperatives to safeguard the confidentiality of data subjects' information. As agencies modernize their SDL implementations and transition to formal privacy accounting of disclosure risk, they will face a number of challenges relating to the decision-making necessary under the triple trade-off.  

None of these challenges are unique to formal privacy. The technical practitioners of SDL who have been responsible for designing and implementing  agencies' SDL mechanisms have, in one way or another, been grappling with each of these challenges for decades. The quantification and composability of formal privacy solutions across the totality of data products to be released, however, brings many of these challenges into sharper focus and makes principled decision-making about how to address them more important. Similarly, the opportunity for transparency afforded by formal privacy, when compared to some traditional SDL approaches, allows decision-making in the context of these challenges to be publicly evaluated and debated in ways that can both inform and complicate those decisions.

\subsection{Identification and prioritization of use cases}
The selection and implementation of any SDL framework and implementing mechanisms will necessarily diminish the potential utility of the resulting data. The choice of which SDL mechanism to use and the selection of certain mechanism parameters over others have direct implications for the utility of the resulting data for different use cases. The only way to maximize utility for every potential use case would be to forgo SDL entirely and publish the data without any confidentiality protections whatsoever \cite{dinur:nissim:2003:10.1145/773153.773173}. Any attempt to reduce disclosure risk, therefore, necessitates that some conceivable use case(s) will be negatively impacted. While official statistics are widely considered a public good, and great importance is often placed on their value for yet-to-be-identified analysis and research questions, principled decision-making about the framework and implementation of SDL requires agencies to identify and prioritize the intended use cases for their data products. Effective prioritization in support of these decisions will yield data products with the highest overall utility and societal value. Failure to identify and prioritize important use cases, however, could result in data products that are uniformly mediocre for any conceivable use case. 

Comprehensive identification and prioritization of use cases, particularly for flagship official statistics products, can be a herculean task. While some important use cases are readily apparent, agencies are often unaware of less visible, but still important, downstream uses of their data products. Nevertheless, maximizing the overall societal value of official statistics requires agencies to make a concerted effort to understand the full spectrum of ways their data are (or will be) used. Identifying these use cases, however, is not enough. Depending on the diversity of these uses, the agency will have to make choices that favor utility for some uses over others. Considering the triple-trade-off, for any given level of confidentiality protection, the quantity of statistics published and the utility of those statistics for differing uses are finite resources that need to be carefully allocated and managed. While it may be difficult, for example, to assess the relative societal value of statistics for city planning versus statistics for public health, these diverse use cases may measure data utility very differently and choices may be necessary as to which is more important when navigating the triple trade-off.  

\subsection{Determining an ``optimal'' privacy-loss budget}
Although statistical confidentiality laws often rely on the simple, but fictional, bifurcation between identifiable and de-identified data, the reality is that the public release of any statistic derived from confidential identifiable data will necessarily carry some disclosure risk \cite{dinur:nissim:2003:10.1145/773153.773173}. Approaching SDL design and implementation from the risk-management perspective assumes that some level of disclosure risk is acceptable in support of the societal value of the official statistics to be published. Compared with the more subjective approach to risk assessment common to many traditional SDL approaches, formal privacy accounting can help support careful and quantifiable assessment and mitigation of disclosure risk within this risk management framework. 

Although formal privacy enables precise quantification of disclosure risk, it does not help answer the question ``How much disclosure risk is acceptable?'' \cite{abowd:schmutte:2019}. Once again, agencies face a difficult decision under the triple trade-off. Greater protection decreases disclosure risk---reducing the potential harm to data subjects and the legal or reputational harm to the agency from a disclosure. But it necessarily reduces the societal value of the data by reducing the quantity and/or utility of the data to be published. Adopting a higher disclosure risk tolerance, on the other hand, can enhance the overall societal value of the data, but increases the potential private harm to individual data subjects. Absent legal or regulatory standards for acceptable disclosure risk and effective use-case prioritization, there is no clear answer to the question of how much risk an agency should accept. Ultimately, the appropriate balance between public benefit and private harm inherent to SDL decision-making is a public policy decision.  Much like other controversial public policy decisions (e.g., ``guns vs. butter'', inflation control vs. full employment) agencies would be well served by augmenting their own expert judgement with input from diverse elements of civil society. 

\subsection{SDL within the workflow of the data life cycle}
The production of official statistics is a complex but integrated workflow of data collection, processing, statistical computation, and dissemination systems. Within this data ecosystem, SDL has often been inserted as a filter between the statistical computation and dissemination stages. While this is often convenient for agencies from the perspective of data product design and operational efficiency, isolation of SDL decision-making and implementation at this point relatively late in the data production cycle can significantly limit the flexibility of SDL methods to optimize data utility for priority use cases within the triple trade-off. If SDL is only considered after data products design specifications have completed (i.e., after table shells and reporting categories are finalized), then SDL practitioners have already lost many of the levers they could use to navigate the privacy-utility-quantity trade-offs that have to be made. Incorporating components of the SDL framework early in the product life cycle, however, may increase the flexibility of SDL solutions to adeptly navigate these trade-offs and optimize the overall societal value and confidentiality of the resulting data. Doing so may also necessitate interruptions, modifications, or even redesign of agencies' finely tuned data production systems. This can be especially problematic for data products that operate on continuous tight production schedules. However, even these systems are regularly re-engineered---for example, all Census Bureau systems must be re-engineered as part of the current information technology initiative \cite{thieme:2022}. Disclosure limitation frameworks that are well-suited to enterprise disclosure risk management could be incorporated into such re-engineering.

\subsection{Communication regarding SDL methods and impact}
Since the ethical practice of statistics obliges statistical agencies to identify and communicate any likely or known limitations of the data they produce that could impact data users' analysis or interpretation of those data \cite{asa:2022}, more transparency in communication about SDL decisions must be part of any modernization effort. Historically, however, statistical agencies have often provided only relatively general or abstract information about the SDL methods they have employed and very little information about the impact of those methods on data utility \cite{abowd:schumtte:2015}. In some cases, such as with suppression or coarsening methods, the data user can discover or infer the broad parameters of the method that was used. For some methods, however, like data swapping and other forms of perturbation, the resulting data may not appear to the average data user as treated with SDL at all. Regardless of the framework and mechanisms selected, the application of SDL always limits on the suitability of the data for particular use cases. Even suppression routines, arguably among the most overtly transparent of all SDL methods, can carry data usage limitations that are not obvious to the data user, such as nonignorable missingness \cite{littlerubin,abowd:schumtte:2015}. Swapping and other traditional so-called perturbative methods, on the other hand, are even more perfidious in this regard, as conventional wisdom has dictated that their parameters (e.g., swap rates) and impact on the resulting data be kept confidential to preserve the integrity of the mechanisms.

Formal privacy methods offer agencies an opportunity to be radically transparent about the design, specifications, and impact of the mechanisms on data utility, but this opportunity for transparency creates its own challenges. The technical complexity of these methods, and their relative novelty, makes effective communication about them to non-technical audiences difficult. Indeed, even for practiced technical users, incorporating such information into their analyses is non-trivial. Effective communication about the limitations of the resulting data will, in many cases, first require effective communication and education about the methods themselves.

\subsection{Governing disclosure risk in the context of sharing administrative and third-party data}
Statistical agencies are always concerned about the relative burden of their information collections, in terms of financial cost and respondent time. As data collection costs increase and survey response rates decline, agencies increasingly turn to administrative and third-party data as a supplement to, or replacement for, their direct data collection activities. Initiatives to support increased evidence-based policy-making in the U.S. have also encouraged expanded use of administrative records and increased data sharing between statistical agencies, although neither the statute (44 U.S. Code §§ 3561-4) nor the accompanying regulations (unfinished) have overcome other legal restrictions on such data sharing (e.g. 13 U.S. Code §§ 8(b), 9) or harmonized the rules regarding data access. The expanded use of administrative and third-party data and of data sharing between agencies poses a significant challenge for SDL generally, and for both formally private and traditional SDL frameworks.  How do you effectively assess and mitigate disclosure risk when multiple agencies are using and publishing data from the same source, even if the decision-making process is extended to include full participation by all contributing agencies? Under rules-based SDL frameworks, the answer was straightforward: the agency supplying the data would specify the SDL mechanisms that had to be applied for publication (e.g., IRS Publication 1075 \cite{IRS:1075}). But traditional rules-based SDL mechanisms are typically unable to quantify incremental disclosure risk across multiple releases---they do not compose. Formal privacy disclosure risk accounting could offer a solution, insofar as it would allow this global privacy-loss accounting across the agencies' data releases, but it would require both agencies to agree to use compatible formally private frameworks and implementing mechanisms. Governance of the overall privacy-loss budget could also pose a challenge, particularly if the agencies have differing perspectives on the relative importance of various data products and/or use cases.

\section{Principles and best practices }
Modernization of SDL for the $21^{st}$ Century requires statistical agencies to transition to formal privacy methods that can quantify and account for incremental disclosure risk across multiple data releases. Failure to do so will make official statistics increasingly vulnerable to known and emerging vectors of attack. But the transition to formal privacy will likely be a lengthy and challenging process for statistical agencies. The U.S. Census Bureau's recent experience transitioning SDL for the Decennial Census of Population and Housing to formal privacy has offered a number of important lessons that can help guide agencies through this transition.

\subsection{Start by acknowledging the reality of disclosure risk}
The inescapable triple trade-off of official statistics means that statistical agencies must make difficult choices about the quantity, utility, and confidentiality of the data they produce. A prerequisite for that decision-making is the acknowledgement that disclosure risk in official statistics is a real threat, that it can only be managed rather than eliminated, and that any mitigation of disclosure risk will necessarily impact the quantity and/or utility of the data that can be publicly released. There will be legitimate debate over how much disclosure risk is acceptable. There will be conflict over which use cases should be prioritized. And there may be differences of opinion over where and how confidentiality protections should be applied. But ignoring the fundamental reality that publicly releasing large quantities of statistics derived from confidential identifiable information is, to a greater or lesser degree, inherently disclosive will only serve to impede principled decision-making about how best to manage disclosure risk and to balance the societal value of official statistics against the potential individual harm from disclosure.  

\subsection{Consider SDL in the broader life cycle of data product releases, rather than piecemeal}
The flexibility of any SDL approach in the context of multiple, successive data releases is inherently path-dependent. Choices made for earlier data products, whether the selection of SDL method to use or the amount of privacy-loss budget to expend, will constrain the flexibility of SDL options for subsequent data products. The adoption of formal privacy disclosure risk accounting will support agencies in assessing and mitigating disclosure risk across multiple data releases, but agencies will need to be mindful of how they manage their privacy-loss budgets across those releases. Researchers often come up with high-value, innovative, and unforeseen uses of the confidential data long after the full suite of agency data products have been released. Supporting those uses should not automatically come at the cost of decreased confidential protection (by exceeding the previously established global privacy-loss budget). Rather, agencies should proactively consider the potential for these unforeseen uses in their overall decision-making under the triple trade-off over the whole life cycle of the data by choosing an appropriate horizon over which the privacy-loss budget applies then reserving some of that budget for unforeseen projects.

\subsection{Begin SDL planning early in the data life cycle}
Small changes in data product specification can yield outsized impacts on the resulting disclosure risk of the data. Reducing (or expanding) reporting categories and removing (or adding) cross-tabulations can greatly affect how easy (or difficult) it will be to mitigate disclosure risk. In 1995, the privacy community began advocating for ``Privacy by Design'' in the context of commercial technology \cite{cavoukian:privacy:2010}. The premise was simple: it is much easier to incorporate effective privacy safeguards early in the product development process than it is to fix privacy problems when they are identified in an already developed product. Similarly, effective management of the privacy/utility/quantity trade-off will be much easier if it is addressed early in the data product life cycle, before final decisions on data product specifications are made.

\subsection{Governance (but not necessarily implementation) of SDL needs to be centralized}
Historically, much of the decision-making regarding SDL for official statistics has been considered the technical purview of the agencies' disclosure review boards (DRBs) and SDL practitioners. The importance of DRBs and highly trained SDL experts cannot be overstated: assessing and mitigating disclosure risk is a highly technical discipline that requires specialized expertise to do effectively. That said, a recurring theme through many of the challenges to modernizing SDL identified above is the public policy nature of decision-making regarding SDL. Decisions about how much disclosure risk is acceptable and the proper balancing of competing data use cases should be made by agency leadership in a coordinated and centralized way, giving clear guidance to the SDL experts for them to implement. This does not mean that all application of SDL need be centralized within an agency---there are many statistical agencies that have multiple SDL experts within different operating units each with specialized knowledge about their respective area's data products---merely that governance of disclosure risk assessment and mitigation should be centralized to ensure that the agency's priorities are being consistently applied across the organization.    

\subsection{Involve (and educate) key internal stakeholders in decision-making}
Because of the triple trade-off, decision-making about SDL is intricately entwined with decision-making about data production and dissemination. As such, any decisions about balancing the societal value of official statistics with the legal and ethical imperative to safeguard data subjects' information should only be made with the full input of diverse stakeholders within the agency. Agencies' policymakers, privacy officials, legal counsel, data management specialists, and data subject matter experts will all have important perspectives on how to balance the countervailing obligations of the agency. Effective dialogue between these stakeholders, to inform agency leadership's decision-making, however, requires that these diverse individuals have a baseline of knowledge and understanding about the nature of disclosure risk assessment, mitigation, and the trade-offs on data availability and utility that SDL implies. 

\subsection{Engage and educate external partners and data users before decisions are made to inform key decision-making}
Official statistics only have value if they are used. Managing disclosure risk in official statistics will, necessarily, impact that value to society. Similarly, the decisions that agencies make regarding the selection and implementation of SDL can have distributional impacts for different groups on the resulting value of---or potential harm from---the publication of data products. Effective decision-making on these issues should be done with as close to complete information about these impacts as is practical. To ensure that these diverse but important viewpoints can be considered when those difficult decisions are being made, agencies should endeavor to consult with a diverse array of their external partners and data users before any SDL decisions are made. Effective dialogue with these external stakeholders, who may not be familiar with the technical domain of disclosure risk assessment and mitigation, may first require some education about the issues involved and how they relate to data availability and utility.

\subsection{Recognize that incremental modernization of disclosure control is OK and may be necessary}
The disclosure risk landscape is always changing, and modernization of SDL should be seen as a journey rather than a destination. Formal privacy solutions offer enormous potential for addressing some of the more vexing vulnerabilities inherent to many traditional approaches to SDL. But formal privacy is also a relatively new discipline, and much research and development will be needed to fully achieve its potential across the diverse universe of statistical data products. As such, agency modernization of SDL should not be considered an ``all or nothing'' endeavor. Sometimes, in the context of production schedules, operational requirements, or resource constraints, it may be necessary to postpone a more complete transformation of an agency's SDL regime in favor of incremental improvements to existing methods. Doing so is not an abdication of the agency's obligation to safeguard data subjects' information in any way, provided the decision to do so is made in a principled way, based on the best available research and internal and external stakeholder perspectives, and with a full understanding of the risks and benefits that it would entail. 

\section*{Acknowledgements and disclaimer}
We thank Gary Benedetto, Monique Eleby and Sallie Keller for helpful comments. The views expressed in this chapter are those of the authors and not the U.S. Census Bureau. This chapter is forthcoming in the CRC Handbook of Formally Private and Synthetic Data Approaches for Statistical Disclosure Control.







\bibliographystyle{plain}
\bibliography{references}

\printindex
\cleardoublepage
\end{document}